\documentclass[final,3p,times]{elsarticle}

\usepackage{amssymb}
\usepackage{hyperref}
\usepackage{breakurl}
\usepackage{comment}

\journal{Carbon}

\begin{document}

\begin{frontmatter}

\title{Electron enrichment of zigzag edges of armchair--oriented graphene nano--ribbons increases
their stability and induces pinning of Fermi level}

\author{E. Louis$^{1,2,*}$, E. San-Fabi\'an$^{2,3}$, G. Chiappe$^{1,2}$, J.A.
Verg\'es$^{2,4}$}

\address{
$^1$Departamento de F\'{\i}sica Aplicada, Universidad de Alicante, 03690
Alicante, Spain\\
$^2$Instituto Universitario de Materiales de Alicante (IUMA) and Unidad Asociada
del Consejo Superior de Investigaciones Cient\'{\i}ficas,
Universidad de Alicante, 03690 Alicante,Spain\\
$^3$ Departamento de Qu\'{i}mica F\'{i}sica, Universidad de Alicante, E-03080
Alicante, Spain. \\
$^4$Departamento de Teor\'{\i}a y Simulaci\'on de Materiales, Instituto de
Ciencia de Materiales de Madrid (CSIC), Cantoblanco, 28049 Madrid, Spain.}
 
\begin{abstract}
Zigzag edges of neutral armchair--oriented Graphene Nano--Ribbons show
states strongly localized at those edges.
They behave as free radicals that can capture electrons during processing,
increasing ribbon's stability. Thus, charging and its consequences should be investigated.
Total energy calculations of finite ribbons using spin--polarized Density Functional
Theory (DFT) show that ribbon's charging is feasible.
Energies for Pariser-Parr-Pople (PPP) model Hamiltonian
are compatible with DFT allowing the study of larger systems.
Results for neutral ribbons indicate: i) the fundamental gap of spin--polarized (non--polarized) solutions is 
larger (smaller) than experimental data, ii) the ground state is spin--polarized,
a characteristic still not observed experimentally. Total energy of GNRs decreases with the number of captured 
electrons reaching a minimum for a number that mainly depends on zigzag--edges size. 
The following changes with respect to neutral GNRs are noted: i) the ground state is not spin--polarized, 
ii) fundamental gap is in-between that of spin--polarized and non--polarized solutions of neutral ribbons,  
iii) while in neutral ribbons valence and conduction band onsets vs. the
fundamental gap, linearly and symmetrically approach mid--gap with slope
0.5, charging induces Fermi level pinning, i.e., the slopes of the valence and conduction bands 
being about 0.1 and 0.9, in agreement with experiment. 
\end{abstract}

\end{frontmatter}

*Corresponding author. E-mail: enrique.louis@ua.es (Enrique Louis) 



\section{Introduction}

The discovery of graphene has triggered a wealth of studies, both at research
and production Institutions, aiming to find out applications for such a novel
material \cite{MTC-4-1,CSR-46-4572,NSR-5-90}. Possible applications cover many
industrial sectors requiring a panoply of properties that may justify referring
to graphene as {\em the miracle material}. Requirements go from being a defect free
well--defined gap material as in most applications related to nano-- and
micro--electronics \cite{NC-11507,AM-28-6222} up to the unavoidable
functionalization of graphene if the aim is to use this material in any kind of
chemical reaction (catalysis, sensors, electrochemistry, ...) \cite{MTC-4-1}. 

\begin{figure}
\includegraphics[angle=0,width=\columnwidth]{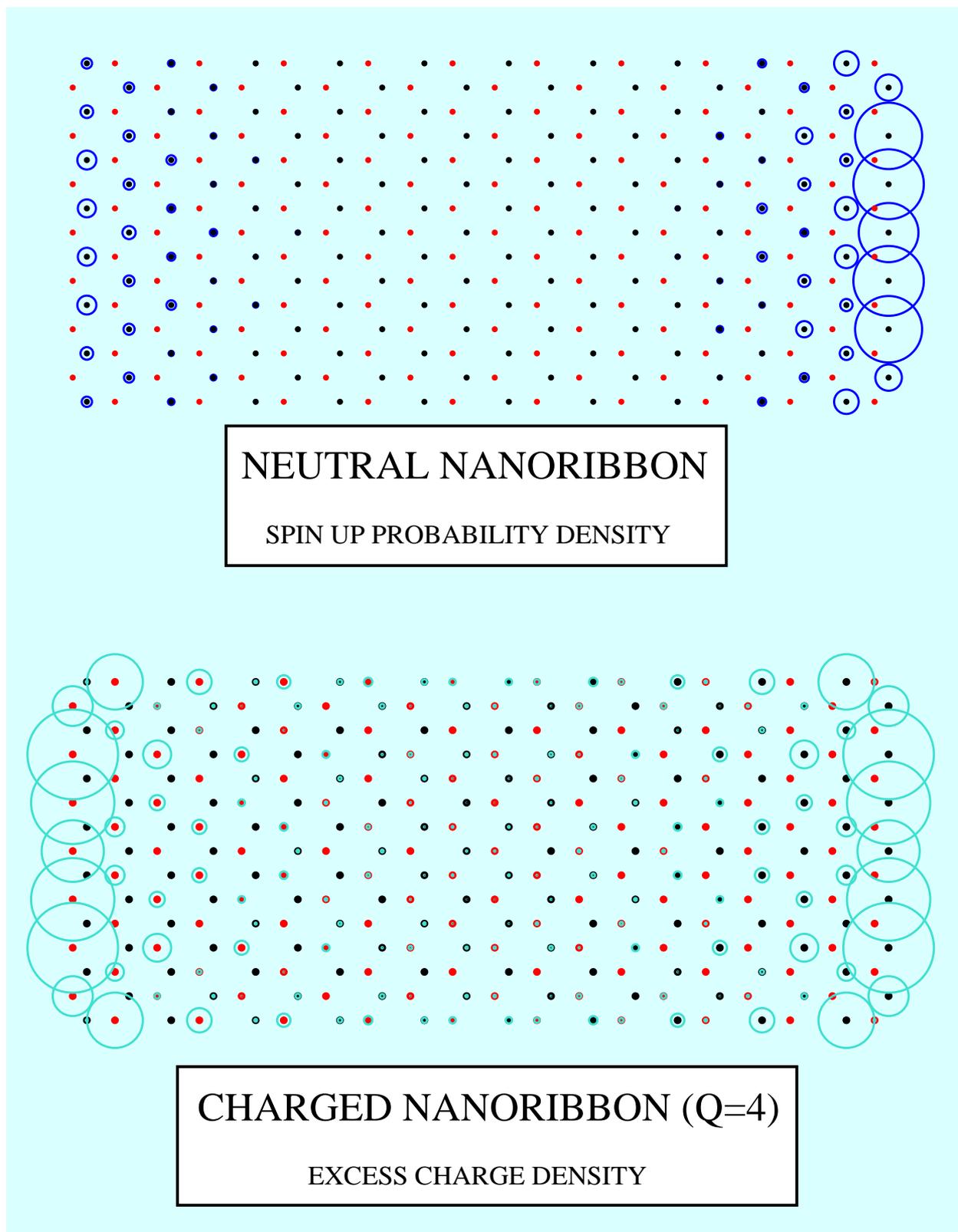}
\caption{(Color online) Upper panel: Spin up probability density of a neutral
\{20,15\}--GNR with equal number of up and down electrons. Spin down
probability density is obtained by a simple right--left swap.
Spin polarization is opposite at the two zigzag edges and,
consequently, breaks spatial inversion symmetry.
Lower panel: charge distribution of the
excess charge of the same ribbon after adding four electrons (Q=4).
Results correspond to UHF solutions of PPP Hamiltonian.}
\label{Sz-q}
\end{figure}
 
Controlled functionalization usually requires starting from defect--free pieces
of graphene. Thus, applications in the two industrial sectors mentioned above
utilize preferably {\em perfect} ribbons. In the last few years bottom--up techniques
have succeeded in producing nearly perfect Graphene Nano--Ribbons (GNRs) very
particularly long rectangular ribbons showing long armchair edges and short
zigzag edges. We will refer to these kind of samples as armchair--oriented
GNRs reserving the name Armchair Graphene Nano--Ribbons (AGNRs) to those infinitely
long ribbons exclusively limited by armchair edges. Bottom--up fabrication
strategies, that rely on the coupling and subsequent cyclo--dehydrogenation of
suitable precursor molecules on a metallic surface (mostly gold
surface)\cite{NC-11507,AM-28-6222,PRB-93-241403,MH-3-186,AJP-69-960,JCTC-11-2003,
C-76-285,ACSN-11-1380,ACSN-7-6123,JACS-135-2060,ACSN-6-6930,NN-7-713,
APL-103-253114} have been developed up to the point that ribbons with
well--defined shape and dimensions, and thus, forbidden gap, can now be produced
in many laboratories \cite{AM-28-6222}. In addition, transferring the GNR by
means of a STM tip to a thin NaCl deposit onto the metallic substrate has
allowed a reliable characterization of the electronic structure of the GNR. 
According to the authors of Ref.\cite{NC-11507} the whole process neither deforms
nor changes the charge--state of the ribbons.  

Finite GNRs show both armchair and zigzag edges,
although the GNRs commonly used in experimental studies are usually much longer
along the armchair direction (i.e., armchair--oriented GNRs as defined previously).
The much shorter zigzag edges of these GNRs host localized electronic
states \cite{ZC-13-200,FC-1-22} that are predicted to be spin polarized 
\cite{PRL-99-186801,RPP-73-056501,PRL-111-085504}. However, spin polarization
at GNRs edges has still never been observed. It is argued that such localized states
may induce defect formation, edge roughness, ..., or interact with the 
substrate and, as a consequence, the study (both experimental and
theoretical) of their intrinsic electronic and magnetic structures is difficult.
As remarked
in \cite{JPCC-119-16991} "the armchair edges and the region surrounding them are
substantially {\it more  aromatic}, compared to zigzag edges". This is probably
the origin of the much less technological difficulties that the fabrication of
armchair--oriented GNRs seems to pose, as compared to
zigzag--oriented GNRs. Our study shows that these molecular orbitals localized at
zigzag edges of neutral GNRs may act as free radicals that can capture
electrons during ribbon processing. When this happens edge reactivity vanishes
and the whole ribbon's stability increases as the behavior of total energy demonstrates.

It is obvious that, although no experimental evidence of charge capture by GNRs
is yet available (see Ref.\cite{NC-11507}), the foregoing remarks make mandatory
the investigation of the charging process and its consequences.
In this work, electron charging of freely suspended {\em perfect} GNRs is
investigated. Calculating the total energy of a ribbon with a given size
(i.e., $\{m,n\}$, $m$ and $n$ being the number of carbon atoms along armchair
and the zigzag directions, respectively) as a function of 
the amount of charge it may capture, we obtain in all cases
a curve with a single minimum that gives the most probable charged state of the 
nano--ribbon. The total energy of small ribbons were firstly calculated both
by means of {\it spin polarized} Density Functional Theory (DFT) and
mean--field UHF solutions of PPP model. Since results provided by both methods
are consistent, the less demanding model (PPP) was subsequently employed.

Let us finally say that thanks to GNRs charging the evolution of levels
around the gap as a function of ribbon size changes completely.
While in neutral ribbons valence and conduction band onsets (HOMO and
LUMO levels) plotted vs. the fundamental gap, linearly and symmetrically
approach the mid-gap with slope 0.5, charging breaks this symmetry inducing
Fermi level pinning as the band gap decreases \cite{ACSN-11-11661}; the slopes
of the valence and conduction bands being close to about 0.1 and 0.9, to be
compared with the experimental data, 0.08 and 0.92, respectively.     

\begin{table}
\begin{center}
\caption{Calculated total energies (all in eV) of small \{{\em m},{\em n}\}--GNRs referred to
the energy of the respective neutral ribbon. Calculations were carried out by
means of either DFT with the functional/basis set B3LYP/6-31+G*-SD,
Unrestricted--Hartree--Fock (UHF) solutions of Pariser--Parr--Pople (PPP) model Hamiltonian
or PM6. Charge at which energy is minimal in each case is shown in bold
characters. Note that DFT energies are systematically lower than those provided by
mean field solutions of PPP Hamiltonian.}
\begin{tabular}{cc|cc|cc|cc}
   &  &   \multicolumn{2}{c}{DFT}  &    \multicolumn{2}{c}{PPP} & \multicolumn{2}{c}{PM6} \\
\cline{3-4}\cline{5-6}\cline{7-8}
       &     Charge  &  {\em m} = 14  &   {\em m} = 20  &   {\em m} = 14  & {\em m} = 20   &  {\em m} = 14   &  {\em m} = 20   \\
\hline
 {\em n} = 7       &2     & {\bf -3.32}   &  {\bf -3.76}   & {\bf -2.82}        &{\bf -3.22}     & {\bf -3.68}   &  {\bf -4.02} \\
                 &4     &  1.92    &  -0.90       &  3.25& 0.61     &  -0.38   & -3.03   \\
                 &6     &  14.63   &  8.52       & 17.90 & 10.91  &  10.61  & 4.71  \\
                 &8     &  32.10   &  23.95       &39.65 &27.85    &  29.13 & 17.91  \\
\hline              
{\em n} = 9     &2     & {\bf -3.39}    &{\bf -3.79} &    {\bf -1.38}      & {\bf -3.02}  & {\bf -2.84}  & -3.13\\
              &4     & 0.18     &  -2.33  &      1.95     &   -0.26  &   -1.47  &  {\bf  -3.30} \\
              &6     & 11.62    & 5.58  &        14.42    &  8.65  & 8.02  & 3.51\\  
              &8     &   28.92    &    19.93   & 34.09 &    23.96    &  23.70  & 15.19\\
\hline              
{\em n} = 15    &2     &  {\bf -3.98}   &   {\bf -4.88}  & {\bf -1.33}   &    -1.33   &  {\bf  -5.92}  &  -6.20  \\
              &4     &  -2.81   &  -4.74   &  -0.81     &     {\bf -2.03}   & -5.48   &  {\bf  -6.62}         \\
              &6     &  4.34    &   0.13    &  7.12     & 3.24   &  0.59  & -3.31  \\
              &8     &  18.15  &  10.26     & 21.91       & 14.45  &  12.19  & 5.12\\
\hline             
\end{tabular}
\label{ene-B3LYP-PPP-PM6}
\end{center}
\end{table}

\begin{figure}
\includegraphics[width=\columnwidth]{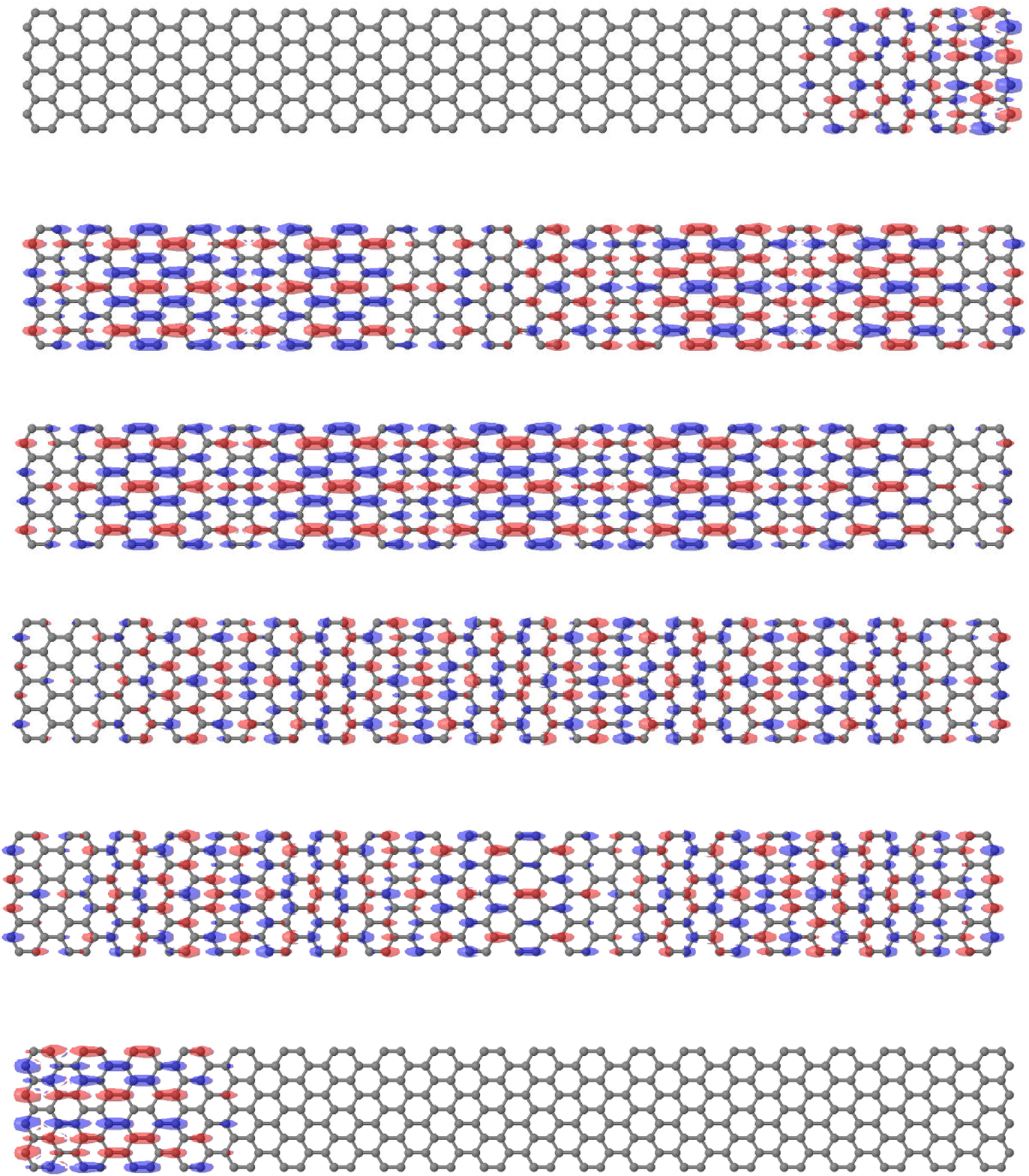}
\caption{(Color online) Six spin up  molecular orbitals  around the Fermi level
of a neutral \{40,9\}-GNR. Uppermost and lowermost orbitals are strongly
localized at the zigzag edges of the ribbon. The fundamental gap lies between
the third and fourth orbitals (starting from the lowermost orbital). The Fermi
level lies midway between those two orbitals, its energy being -3.456 eV. 
Orbital energies, as calculated by means of UHF-PPP, are: -5.68, -5.62, -5.477,
-1.453, -1.31 and -1.25 eV. Spin down orbitals are obtained from the spin up
orbitals by right-left swapping.}
\label{PDF-EF-States-9AGNR-Q0}
\end{figure}

\begin{figure}
\includegraphics[width=\columnwidth]{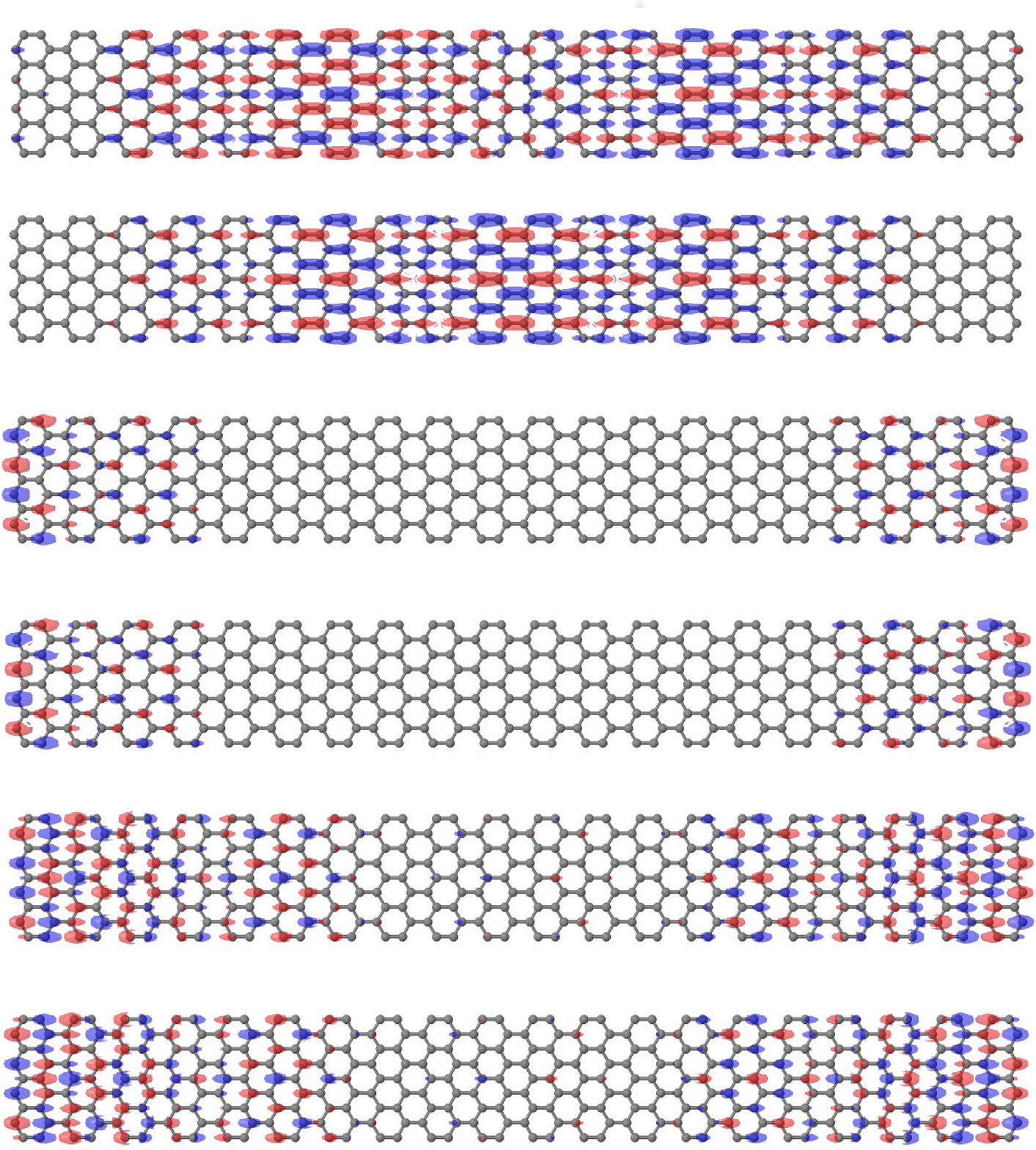}
\caption{(Color online) Six spin up  molecular orbitals around the Fermi level
of a \{40,9\}-GNR charged with two electrons (Q=2). The fundamental gap lies between the
4th the 5th orbitals (starting from the lowermost orbital). The Fermi level lies
midway between those two orbitals, its energy being -1.27 eV. Orbital
energies, as calculated by means of UHF-PPP, are: -3.73, -3.73, -2.14, -2.14,
-0.41 and -0.067 eV. Both $S_z$ and $S^2$ are null (non--magnetic
configuration).}
\label{PDF-EF-States-9AGNR-Q2}
\end{figure}

\section{Methods and numerical procedures}

Calculations have been carried out on freely suspended $\{m,n\}$-GNRs,
being $m$ the number of carbon atoms along the armchair direction and $n$
the number corresponding to the zigzag edge.
Since most of our ribbons show $m > n$, we will associated $m$ to their
lengths and $n$ to their width. Actual size is given by:
$$
{\rm length} = (\frac{3}{2} m - 1) \times l ~~~~~~;~~~~~~
{\rm width} = (n -1) \frac{\sqrt {3}}{2} \times l   ~~~,
$$
being $l$ a typical C-C bond length that we take as 1.41 $\AA$.
The total number of carbon atoms present in the GNR is $N_C= m\times n$.
The chosen ribbons widths were $n$ = 7, 9, 13, 15 and 19, while the largest
length was $m$ = 100 that corresponds approximately to 210.1 $\AA$ (only for the narrower
ribbons). Dangling bonds of all
carbon atoms at the ribbon edges have been bonded to hydrogen so that all
carbon atoms have a single non--saturated $\pi$-like orbital.
Largest ribbons were formed by around 1500 atoms. Extra
electrons were always added in pairs to preclude spin polarized ground states.
Actually, we carried out calculations for Q = 2, 4, 6 and 8.
Calculations were
carried out by means of two semiempirical methods, namely, PPP and PM6.
For the smallest systems spin polarized DFT calculations have also
been carried out.

{\em Ab initio} calculations were done with the Gaussian09 package \cite{G09}.
As regards the exchange-correlation functional, the widely
accepted B3LYP\cite{PRA-38-3098,PRB-37-785,JCP-98-5648,JPC-98-11623} functional
in combination with the 6-31+G*\cite{TCA-28-213,JCC-4-294} basis set has been
used. It incorporates diffuse functions around carbon atoms to get a better
description of delocalized orbitals of charged systems.
Geometries of all ribbons have been optimized.

PM6 \cite{JMM-13-1173} is the latest parametrization of the NDDO method
\cite{JACS-99-4899}, which has 
been modified adopting  Voityuk's core-core diatomic interaction parameters
together with the method of parameter optimization.
Parametrization including  70 elements and the  heats of formation for a subset
of 1,373 compounds (with only H, C, N, O, F, P, S, Cl and Br elements)
give an average unsigned error (4.4 kcal/mol) better than alternative
semiempirical methods, such as RM1(5.0), AM1(10.0) and PM3(6.3), or the
HF/6-31G*(7.4) and B3LYP/6-31G*(5.2). The geometry of all systems was also
optimized.

Finally, PPP Hamiltonian was solved within Unrestricted Hartree-Fock
approximation using parameters fitted to the electronic structure of small
PAH (see \cite{PRB-81-085120} and \cite{JPCM-27-463001}). Note that the fact that
this Hamiltonian includes a single orbital per site should not be a major
drawback as in graphene all carbon atoms have a single non--saturated $\pi$--like
orbital.

\section{Results and Discussion}

Probably forced by the lack of experimental confirmation of the existence of
magnetism in the zigzag edges of GNRs, other theoretical
studies of GNR have discarded unrestricted spin polarized solutions from the
very beginning \cite{C-116-422,PRB-98-155415}. This is so despite of the fact
that mono--determinantal unrestricted solutions are spin polarized no matter the
chosen framework, model Hamiltonians solved within Unrestricted Hartree-Fock, or
the most sophisticated {\it ab initio} DFT and GW approaches
\cite{NC-11507,PRL-99-186801,RPP-73-056501,PRL-111-085504}. In the following we
discuss an alternative way to reconcile the experimental evidence with
calculations. 

\subsection{Localized states at the zigzag edges }

As pointed out in the Introduction, the zigzag edges of neutral GNRs host
unpaired electrons described by localized states that act as radical centers
ready to capture charge. This is clearly illustrated in Fig.\ref{Sz-q} which
shows that the staggered magnetization in a neutral \{20,15\}-GNR is mostly
localized at the two zigzag edges of the ribbon. In the Figure only the up
component of the spin is shown, the down component is derived from it by
exchanging zigzag sides. Spin polarization illustrates the breaking of the
specular symmetry with respect to the plane perpendicular to the long axis
of the ribbon that splits it in two identical parts. 
If electrons are incorporated to the ribbon, no matter how
they were captured, the ribbon ground state becomes non--magnetic
and the original symmetry of the ribbon becomes restored
(see lower panel of Fig.\ref{Sz-q}).
Both empty and occupied localized states lie around the Fermi level. Changes
in the total number of electrons have therefore important consequences on the
electronic structure around the gap. Let us show two examples of the kind
of states involved in the chemical deactivation via electron enrichment
of a typical GNR.

 
\begin{figure}
\includegraphics[angle=270,width=\columnwidth]{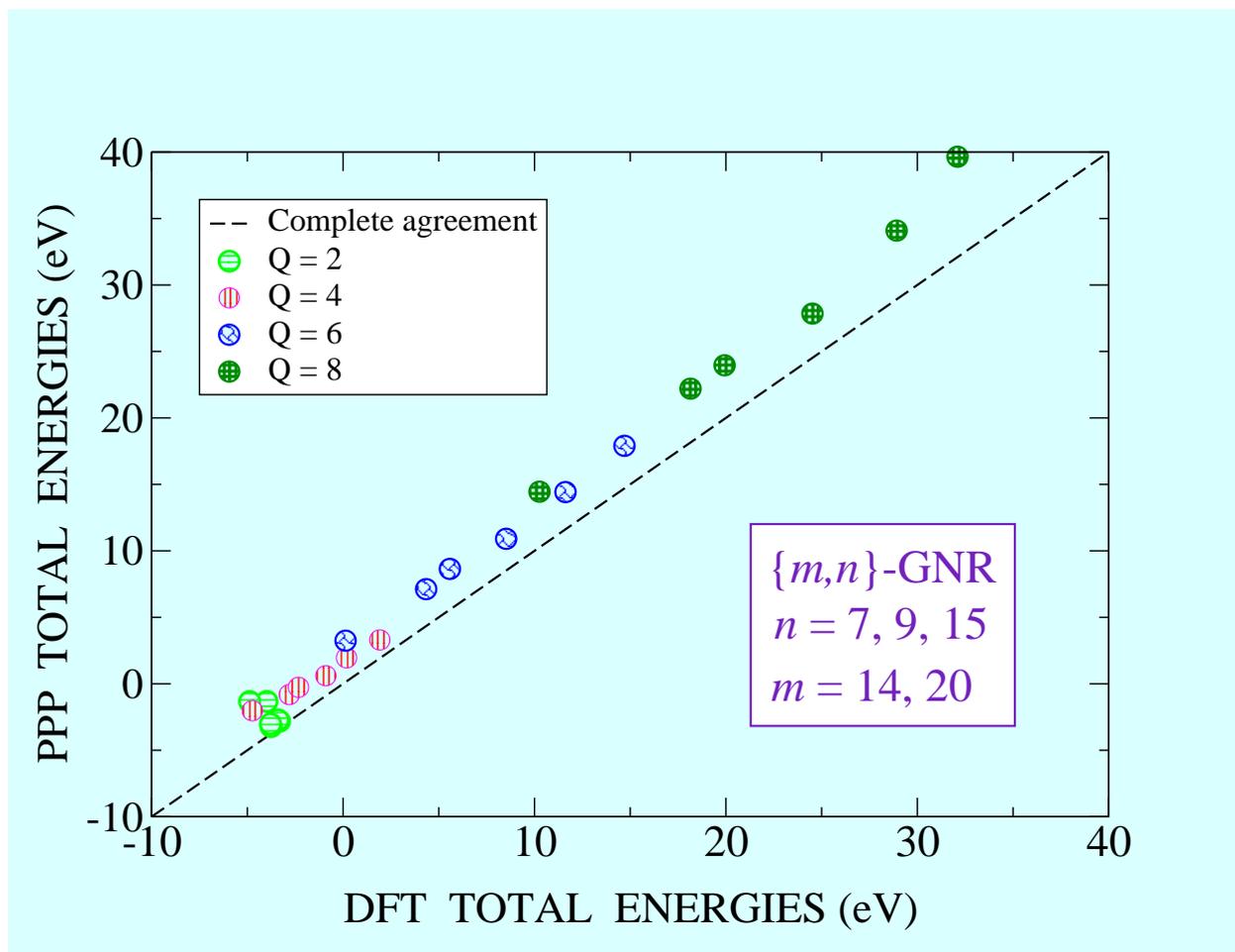}
\caption{(Color online) Energies of charged $\{m,n\}$-GNRs referred to that of
the respective neutral ribbons, calculated by means of HF-PPP versus those
obtained within the DFT framework and the combination functional/wavefunction
set  B3LYP/6-31+G*-SD. The results correspond to the ribbons included in
Table \ref{ene-B3LYP-PPP-PM6}.}
\label{DFT-vs-PPP}
\end{figure}

The characteristics of neutral spin polarized GNRs are
illustrated by the spin densities of the six molecular orbitals around the
Fermi level (see Fig. \ref{PDF-EF-States-9AGNR-Q0}) in a neutral
\{40,9\}-GNR. First and sixth orbitals (starting from the lowermost level)
are strongly localized at the zigzag edges of the ribbon. The Fermi level lies
between the 3rd and fourth orbitals, its energy being -3.456 eV. Molecular
orbital energies calculated by means of UHF-PPP are: -5.68, -5.62, -5.477,
-1.453, -1.31 and -1.25 eV. As already remarked, spin down orbitals are obtained
from the spin up orbitals by exchanging sides.
As a final result, antiferromagnetic correlations between zigzag edges exist.
Instead, extended orbitals are either symmetric or antisymmetric respect the
specular plane mentioned in the previous paragraph (see 2nd up to 5th orbitals
in Fig.\ref{PDF-EF-States-9AGNR-Q0}).

The recovery of the full symmetry induced by charging the ribbons
is nicely illustrated in Fig. \ref{PDF-EF-States-9AGNR-Q2}. This Figure shows
the spin densities of six spin up molecular orbitals around the Fermi level of
a \{40,9\}-GNR charged with Q = 2.
Orbital energies, as calculated by means of UHF-PPP, are: -3.73, -3.73, -2.14,
-2.14, -0.41 and -0.067 eV.
Third and fourth orbitals are strongly localized at the zigzag edges
of the ribbon and fully occupied since the Fermi level lies between the 4th
and fifth orbitals. 
The original symmetry of the graphene lattice becomes {\em all} fully
restored because {\em all} states no matter extended or localized are now
symmetric or antisymmetric. As a consequence,
both $S_z$ and $S^2$ are null (non-magnetic configuration).

\subsection{Charging stabilizes GNRs}

At this point the issue is whether charging reduces the nano-ribbon total 
energy and, thus, increases its stability. Calculated total energies (all in
eV) of small $\{m,n\}$-GNRs by means of DFT, UHF-PPP and PM6, are reported in
Tables \ref{ene-B3LYP-PPP-PM6} and \ref{ene-PPP}.
As shown in the Tables, whenever the excess
charge is not too high, charged ribbons are more stable than neutral ribbons.
This result, which is the core of the present work, holds for the three
theoretical methods used here. In addition it is noted that DFT energies are
always smaller than those obtained with the UHF solution of the model PPP
Hamiltonian. Instead, no conclusion can be derived from a comparison of DFT and
MP6 results. Anyway, the main result is very robust as the three theoretical
methods used here always agree at a qualitative  and, in some cases, even at a
semi--quantitative level: the total energy of GNR diminishes upon ribbon's
charging reaching a minimum at a charging level that depends on ribbon's length
and width. Thereafter, the energy increases monotonically.

\begin{table}
\begin{center}
\caption{Calculated total energies (in eV) of charged \{$m$,$n$\}-GNR for
$m$ = 20, 40, 60, 80 and 100 (corresponding lengths 40.9, 83.2, 125.5, 167.8 and
210.1 $\AA$, approximately) and $n$ = 7, 9, 13, 15 and 19 (for the three largest $n$ results for $m$=100 are not shown). Charge at which energy is minimal in each case is shown in bold
characters. Only results
obtained by the less demanding UHF--PPP scheme are presented.
Energies of charged ribbons referred to the energy of the respective neutral ribbon.}  
\begin{tabular}{lc|ccccc}
\multicolumn{2}{c}{}  &  \multicolumn{5}{c}{}  \\
\hline             
       &Charge  & $m$ = 20   &   $m$ = 40    &   $m$ = 60     &  $m$ = 80     &  $m$ = 100    \\
\hline
   $n$ = 7    &2     & {\bf -3.22} &  {\bf -3.59} & -3.67       & -3.71       & -3.72            \\
       &4     &   0.61      & -2.73        & {\bf -3.79} & {\bf  -4.33} & {\bf -4.69}   \\
       &6     &  10.91      &   1.0        & -1.5       &  -3.06      &     -4.0   \\
       &8     &  27.85      & 9.66         &   3.55       &   0.21       &    -1.74       \\
\hline
   $n$ = 9    &2     & {\bf -3.02}& {\bf -3.33} & -3.42      &  -3.41    &3.45               \\
       &4     & -0.26      & -3.25       & {\bf -4.26} & {\bf -4.74}   & {\bf -4.78}           \\
       &6     &  8.65      & 0.40        &  -2.71      &   -4.21     &      -4.25 \\
       &8     & 23.96      & 7.96        &  -1.63      &    -1.62     &    -1.67   \\
\hline
   $n$ = 13    &2     &   {\bf -3.8}   & -1.87      &   -1.93     &   -1.96         &             \\
       &4     & -2.18& {\bf -4.05} &   -4.59     &   -4.81                 \\
       &6     &    4.22   & -2.39       & {\bf -4.97} & {\bf -5.95}              \\
       &8     &    16.71   &  3.20       &   -2.16     &   -4.93                \\
\hline
   $n$ = 15    &2     &  {\bf -3.51}     &-1.51        &   -1.57     &     -1.84                \\
       &4     &  -2.17& {\bf -3.49} &   -4.38     &     -4.38                         \\
       &6     &   3.24      & -2.41        & {\bf -4.53}  & {\bf -5.57}            \\
       &8     &   14.45     &  2.52        & -2.35        &      -4.91           \\
\hline
   $n$ = 19    &2     &  {\bf -4.07}  & -2.43        &    -2.56    &  -2.62           \\
       &4     & {\bf -4.07}     & {\bf -4.04}  &     -2.56   &  -5.59             \\
       &6     &  1.04       & -3.08        & {\bf -5.36}  &  -6.81             \\
       &8     & 10.09       &  0.49        &      -3.62  & {\bf -6.81}            \\
\hline             
\end{tabular}
\label{ene-PPP}
\end{center}
\end{table}

\begin{figure}
\begin{tabular}{c}
\includegraphics[angle=270,width=0.55\columnwidth]{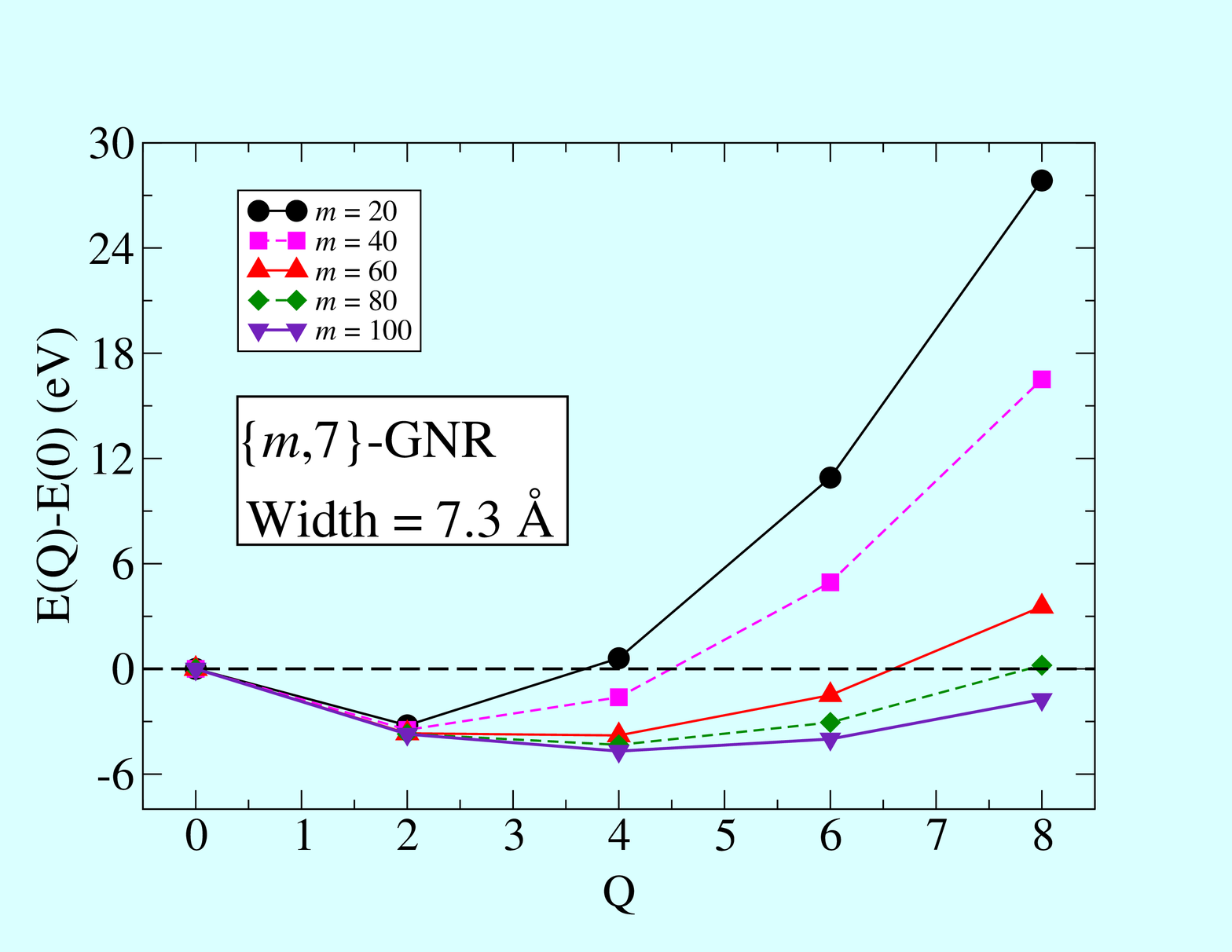} \\
\\
\includegraphics[angle=270,width=0.55\columnwidth]{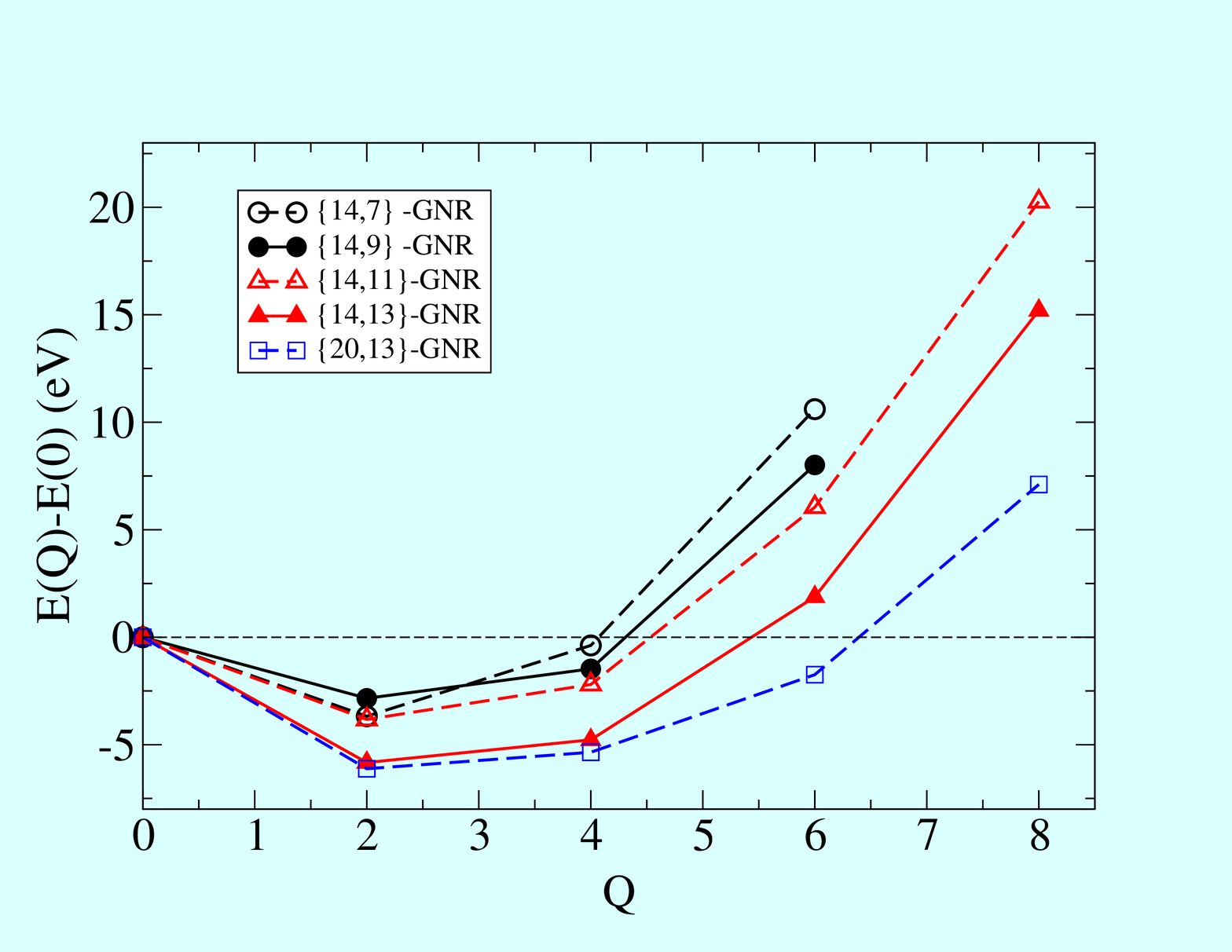} \\
\\
\includegraphics[angle=270,width=0.55\columnwidth]{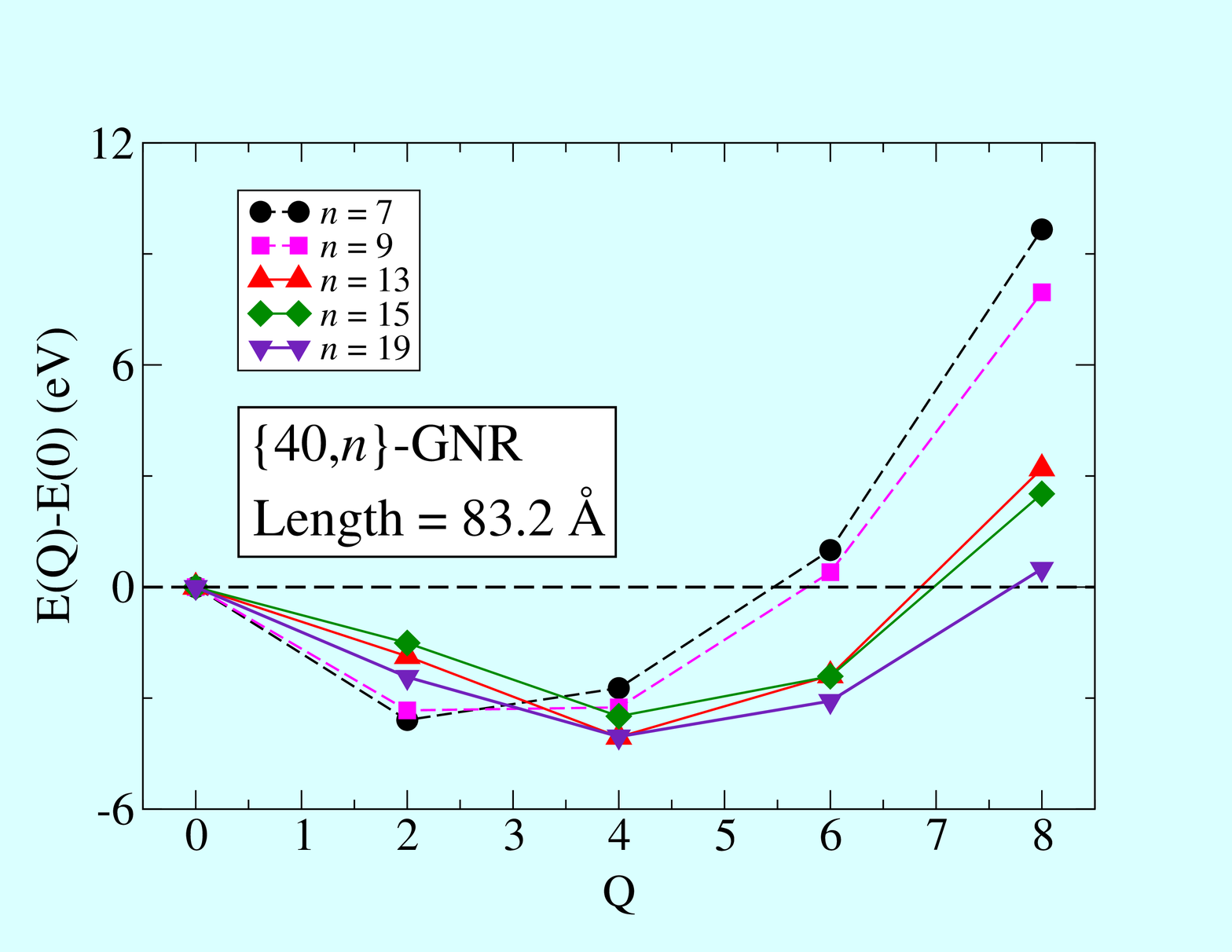} \\
\end{tabular}
\caption{(Color online) Total energies (all in eV) of  several charged GNR as a
function of extra electrons Q. Upper: \{$m$,7\}-GNR and five lengths $m$=20, 40, 60, 80 and
100 (actual lengths excluding H atoms are approximately, 40.9, 83.2, 125.5, 167.8 and
210.1 $\AA$). Middle: Results for  $n$=7, 9, 11 and 13 of two  lengths  versus
the excess  charge. Lower:  Results for GNRs of a fixed length $m$=40 and five
widths $n$=7, 9, 13, 15 and 19. Total energies, referred to
that of the corresponding neutral ribbon, were calculated with PPP (upper and
lower) or PM6 (middle). This Figure clearly illustrates the dependence  of
charging and its effects on ribbons width and length.}
\label{E-vs-Q}
\end{figure} 

Aiming to  illustrate in a pictorial way the agreement of DFT  and HF-PPP
results, Figure \ref{DFT-vs-PPP} depicts the energies of charged $\{m,n\}$-GNRs
referred to that of the respective neutral ribbons. The results correspond to
ribbons charged with Q = 2, 4, 6, 8 and sizes $m$ = 14 and 20 and $n$ =  7, 9 and
15. (see Table \ref{ene-B3LYP-PPP-PM6}). The straight line corresponds to
complete agreement, that is, all DFT energies coinciding with all UHF-PPP
energies. Note that the numerical data are not far from this {\it ideal} line
validating the use of  HF-PPP for the calculation of total energies, fundamental
gaps, ... etc.
Once we have confirmed the consistency between results provided by different methods,
much larger systems have been exclusively calculated using the UHF-PPP scheme.
Table \ref{ene-PPP} provides the corresponding results. Charge at which
energy is minimal in each case is shown in bold characters in both Tables
\ref{ene-B3LYP-PPP-PM6} and \ref{ene-PPP}.

Numerical data compiled in both Tables are better analyzed after a graphical
representation that is given by Figure \ref{E-vs-Q}.
In the upper panel results for \{$m$,7\}-GNR and five lengths $m$ = 20, 40, 60, 80 and
100 have been represented. It can be observed that the minimum shifts from
Q = 2 to Q = 4 for the larger GNRs. Although results for even larger ribbons are
difficult to obtain, we believe that four is the maximum number of electrons that a
$n$ = 7 GNR admits. Notice that a non--completely screened Coulomb interaction between
electrons at opposite zigzag edges makes the convergence of energy vs. length (that is, $m$) very slow.
The middle panel depicts some results obtained with the semiempirical PM6 method.
Energy minima for {\em short} GNRs are obtained for Q =2. Nevertheless,
the total number of electrons stabilizing a zigzag edge should increase with its size.
This is confirmed by the lower panel showing a stabilization with four electrons
for  \{40,13\},  \{40,15\} and \{40,19\} ribbons.
Summarizing, the number of electrons stabilizing a zigzag edge of a given width
goes to a constant value for very large GNRs. On the other hand, the number of
extra charge that admits a zigzag edge is more or less proportional to its
width.

\begin{figure}
\begin{tabular}{c}
\includegraphics[angle=270,width=0.87\columnwidth]{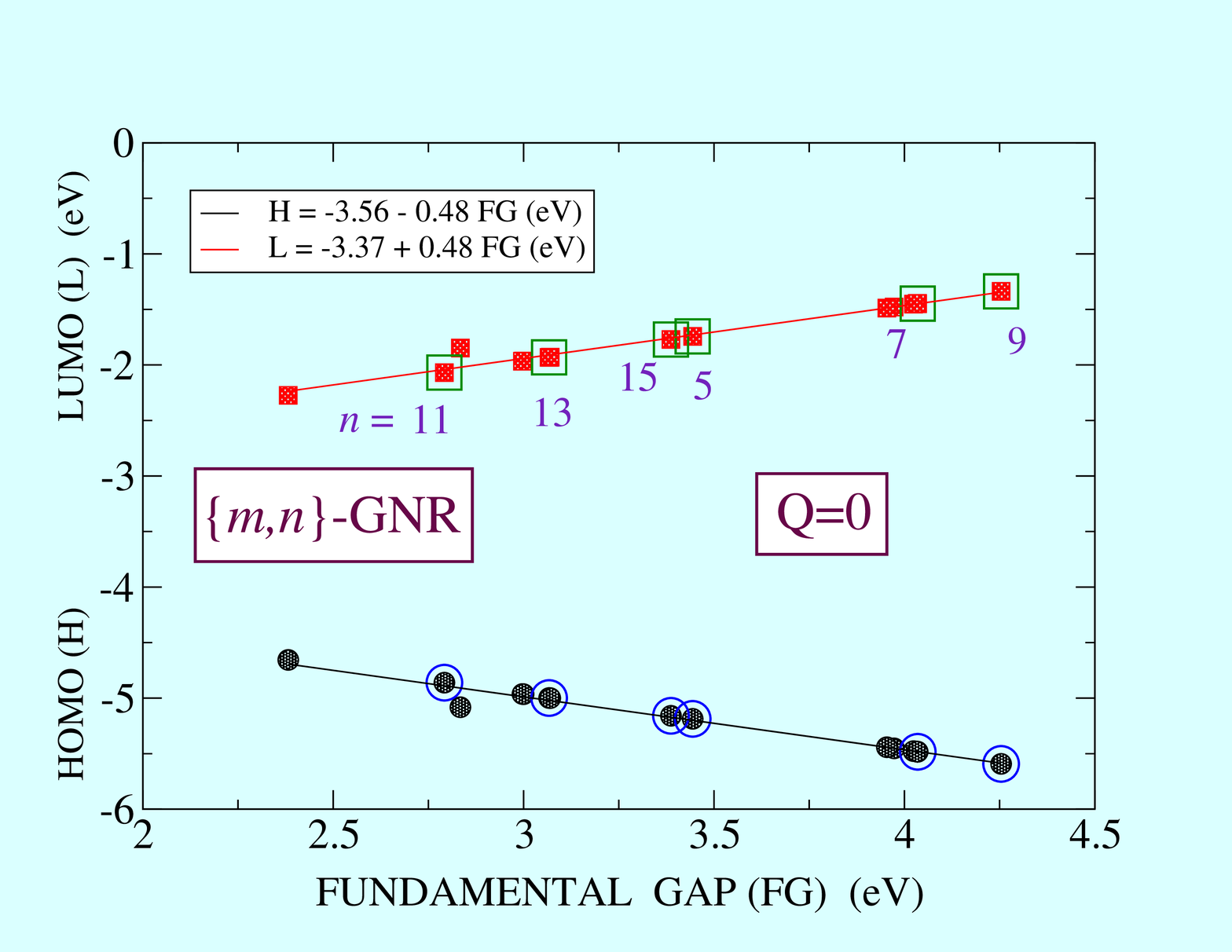}  \\
\\
\includegraphics[angle=270,width=0.87\columnwidth]{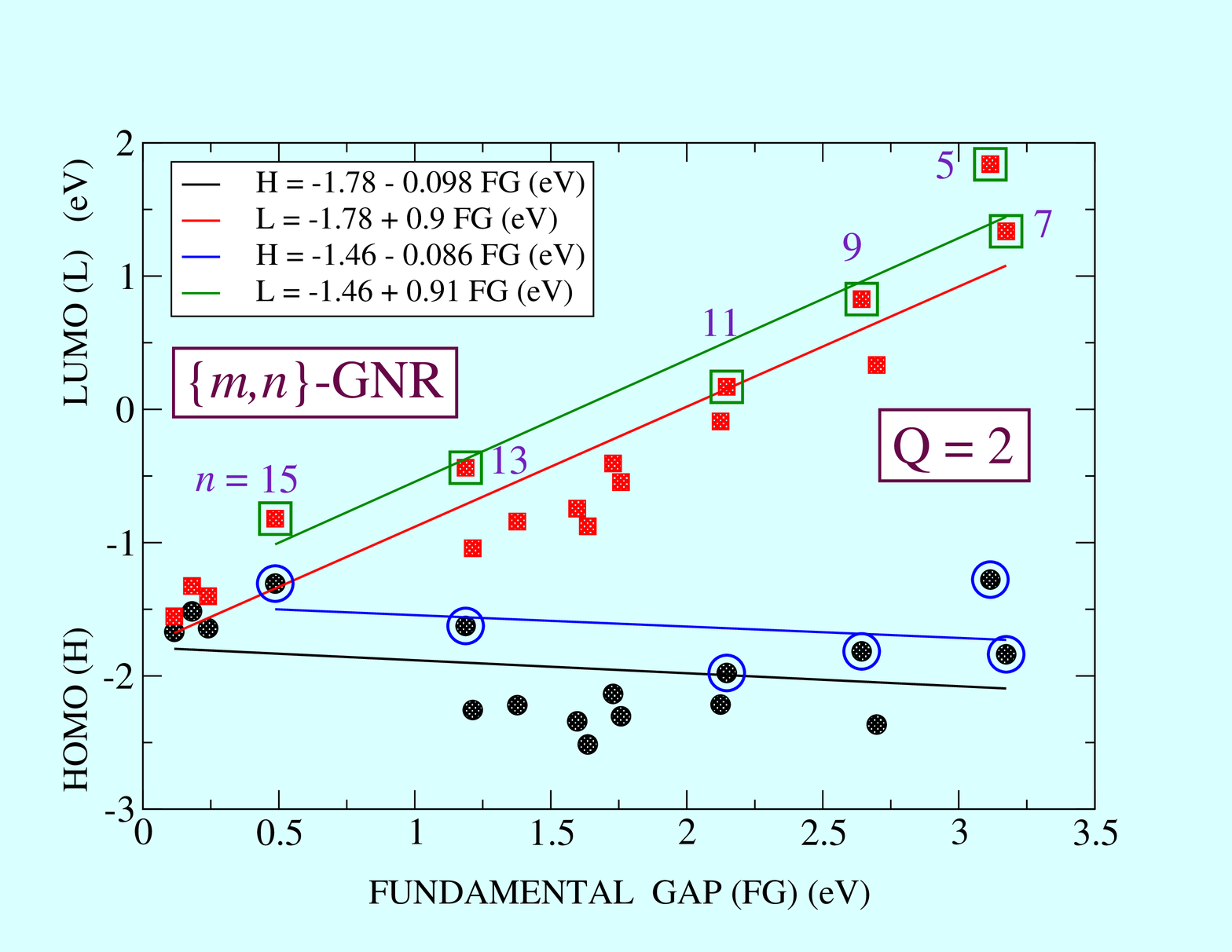}  \\
\end{tabular}
\caption{(Color online) HOMO (circles) and LUMO (squares) energy levels versus 
fundamental gap for graphene nanoribbons of widths $n$=5, 7 and 9 and lengths of
$m$=20, 40, 60 and 80 (that is, 40.9, 83.2, 125.5 and 167.8 \AA, ~approximately),
and of width $n$= 11, 13 and 15 and lengths $m$=20 and 40 (about 40.9 and 83.2 \AA).
Empty symbols correspond to $m$=20 ribbons.
Insets give the straight lines that better fit our numerical results.
Spin polarized solutions were obtained for neutral ribbons (Q=0) within UHF--PPP
scheme (upper panel) while non--magnetic solutions were got after the
addition of two electrons to the same set of GNRs (Q=2, lower panel).}
\label{VB-CB-FG}
\end{figure}

\subsection{Fundamental gap of neutral and charged GNRs versus ribbon width and
length} 
 
The upper panel of Figure \ref{VB-CB-FG} depicts results for the HOMO
(circles) and LUMO (squares) energy levels versus the fundamental gap (FG) for neutral
GNRs of several widths and lengths. Results, obtained with HF-PPP, correspond
to widths of $n$ = 5, 7 and 9 and lengths of $m$ = 20, 40, 60 and 80, and 
widths of $n$ = 11, 13, 15 and 19 and lengths of $m$ = 20 and 40. Empty symbols
correspond to ribbons of fixed length ($m$ = 20). Both HOMO and LUMO levels show a
monotonic dependence on ribbon's FG and a very irregular dependence on  
ribbon's width. The straight lines fitted to the numerical results are shown in
the inset. The slopes are very close to S = $\pm$ 0.5 (valence and conduction band,
respectively), as expected in a system with electron-hole symmetry
\cite{ACSN-11-11661}. 

\begin{figure}
\includegraphics[angle=270,width=\columnwidth]{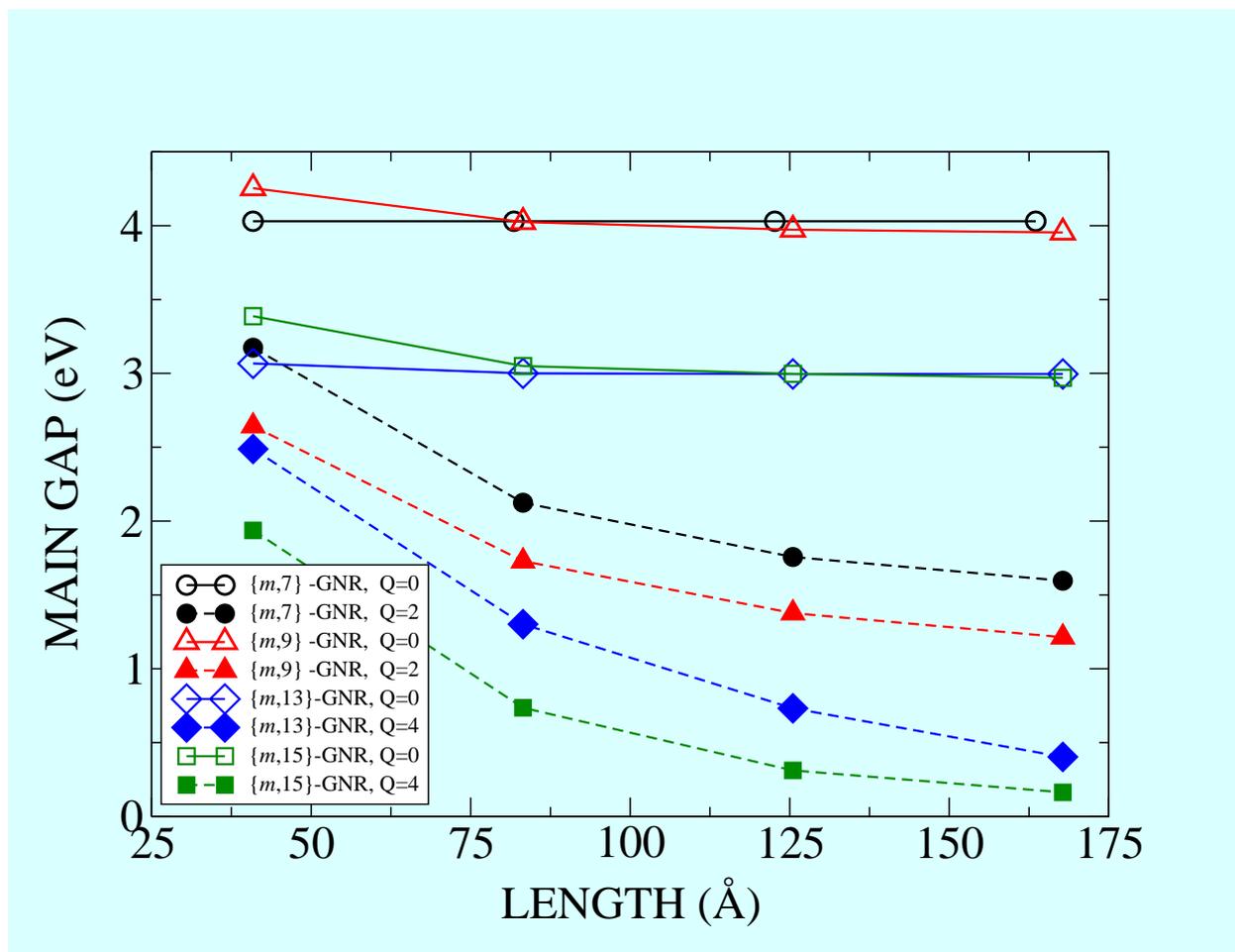}
\caption{(Color online) Fundamental gap for neutral and charged
armchair--oriented GNRs of several widths ($n$ = 7, 9, 13 and 15) versus
ribbon's length ($m$=20, 40, 60 and 80). Results correspond to UHF--PPP scheme.
Neutral GNRs show a gap that varies irregularly with the ribbon's width
remaining constant beyond a length of approximately 80 $\AA$.
Meanwhile, in charged ribbons the gap decreases monotonically with both $n$ and $m$.
}
\label{FG-vs-W-Q-L}
\end{figure}

Recent experiments \cite{ACSN-11-11661} on armchair oriented GNRs of several widths
deposited on
Au(1,1,1) surface report results at variance with the symmetric case illustrated
in the upper panel of Figure \ref{VB-CB-FG}. The authors presented results for
the valence and conduction band onsets versus the fundamental gap (FG) that for
large gap values (FG $>$ 1.7 eV) vary linearly with ribbon's FG with slopes of
approximately $\pm$ 0.5, as found in a symmetric system.
However, for FG $<$ 1.7 eV , the  valence
band shows Fermi level pinning as the fundamental band gap decreases.
The slope of the fitted straight line being slightly smaller than 0.1.
At the same time, the fundamental gap monotonically diminishes with both
$n$ and $m$ (see Fig.5 of Ref.\cite{ACSN-11-11661} and the
dashed curves of our Fig.\ref{FG-vs-W-Q-L}).
We have been able to obtain here similar results for freely suspended
ribbons as long as they are charged with two or more electrons. 
Our results are depicted in the lower panel of Figure \ref{VB-CB-FG}.
They correspond to the same set of ribbons that was employed for the neutral case.
The straight lines fitted to the numerical results are shown in the
inset. When all numerical results are included in the fittings, the slopes are
-0.098 and 0.9 for HOMO and LUMO levels, respectively, not far from the results
reported in Ref. \cite{ACSN-11-11661}  namely, -0.08 $\pm$ 0.05 and 0.92 $\pm$
0.05. Results do not change much if fittings are carried out including only data
for ribbons of length 40.9 \AA. These slopes were obtained from fittings of data
with FG $<$ 3.2 eV, higher than the experimental value of 1.7 eV, a discrepancy
likely due to the mean field approach adopted here. On the other hand, the
results shown in the lower panel of Fig. \ref{VB-CB-FG} reveal a monotonic
dependence of the two band onsets on the ribbon's width, as opposed to what we
obtained for neutral ribbons. This is further illustrated by Fig.
\ref{FG-vs-W-Q-L} which shows that the main gap of charged GNRs
smoothly depends on both length and width.

As a summary of this paragraph we can say
that a wide gap in neutral armchair oriented GNRs is only obtained if a
spin polarized formalism allowing spin symmetry breaking is used, something
not yet fully supported by experiments. Instead, gaps in non--polarized ribbons
are very small no matter which method is used in the calculations. The present
results indicate that charging has several appealing effects: i) the ground
state of the charged ribbons is non--polarized, ii) gaps are substantially
smaller than those found in spin polarized solutions, although are larger than
the experimental ones, and iii) valence and conduction bands are no longer
symmetrically placed with respect to the Fermi level. We believe that these
results are enough to justify the following of the route initiated in this work
as at present there is no alternative theory that can explain the experimental
results of Ref. \cite{ACSN-11-11661}.

\section{Concluding Remarks} 

Zigzag edges of graphene nanoribbons, in particular armchair oriented GNRs, host localized
electronic states of opposite spin at both ribbon edges. Consequently, the most
sophisticated quantum chemistry tools now available, predict a spin polarized
ground state. However, spin polarized edge states have not yet been observed
experimentally. This dramatic disagreement is usually ascribed to the high
susceptibility of these states to develop edge defects and roughness
or to some indefinite interaction with the supporting substrate.
In this work, taking for granted the
predicted spin polarized character of zigzag edges in GNRs and the reliability
of the experimental studies that aiming to find any traces of spin polarization
have failed, we undertook the task of seeking alternative effects that may
obscure or eliminate that spin polarization. Taking note of the fact that some
of the so called {\it edge states} may be empty, we consider the possibility
that adding  extra charge to the ribbons might contribute to increase their
stability. Our work, then, was focused on this issue. All calculations, no
matter which method we used, gave a curve of total energy versus extra charge
that showed a minimum at a value of the charge that varied with ribbon's width
and length. Width dependence of the fundamental gap reveals valence bands which
show Fermi level pinning as the fundamental band gap decreases, in agreement
with recent experimental data and at variance with the symmetry characteristic
of neutral ribbons. It is pertinent to remark that it would be interesting to
investigate the conductance of charged ribbons along the lines, for instance, of
a recently published work \cite{PRB-98-155415}. As regards the actual values of
the forbidden gap it is noted that, in neutral ribbons the gap of spin polarized
gaps is larger than what has been observed up to now, the opposite occurs in
non-polarized. Instead, charged ribbons show a forbidden gap
in-between those two, closer to the experimental data.  

\section*{Acknowledgments}

This work has been partially supported by the Spanish ''Ministerio de Ciencia,
Innovaci\'on y Universidades'' (Grants FIS2015-64222-C2-1-P,
FIS2015-64222-C2-2-P, MAT2016-77742-C2-2-P and AYA2015-66899-C2-2-P), and the
Universidad de Alicante
is gratefully acknowledged.





\end{document}